\documentclass[journal]{IEEEtran}
\usepackage{graphicx}
\usepackage{a4}
\usepackage{fullpage}
\usepackage{epsfig}
\usepackage{epsf}
\usepackage{graphics}
\usepackage{amsmath}
\usepackage{tabls}
\usepackage{amssymb}
\usepackage{verbatim}
\usepackage[latin1]{inputenc}
\usepackage{subfigure} 
\usepackage{fancyhdr}
\usepackage{cite}
\usepackage{ulem}
\usepackage{subfigure}
\usepackage{lipsum}
\usepackage{algorithm}
\usepackage{algorithmic}
\usepackage[space]{grffile}
\usepackage{array}
\usepackage{xcolor}
\usepackage{mathtools}

\begin{document}
\title{Multi--band programmable gain Raman amplifier}

\author{Uiara~Celine~de~Moura,
        Md Asif Iqbal,
        Morteza~Kamalian,
        Lukasz~Krzczanowicz,
        Francesco~Da~Ros,
        Ann~Margareth~Rosa~Brusin,
        Andrea~Carena,
        Wladek~Forysiak,
        Sergei~Turitsyn,
        and~Darko~Zibar
\thanks{U.~C.~de~Moura, F.~Da~Ros and D.~Zibar are with DTU~Fotonik, Department of Photonics Engineering, Technical University of Denmark, DK-2800, Kgs. Lyngby, Denmark (e-mail: uiamo@fotonik.dtu.dk).}
\thanks{Md A. Iqbal, M.~Kamalian, L.~Krzczanowicz, W.~Forysiak and S.~Turitsyn are with Aston Institute of Photonic Technologies (AIPT), Aston University, Birmingham, B4 7ET, United Kingdom.}
\thanks{A.~M.~Rosa~Brusin and A.~Carena are with Dipartimento di Elettronica e Telecomunicazioni (DET), Politecnico di Torino, Corso Duca degli Abruzzi, 24 - 10129, Torino, Italy.}
\thanks{Manuscript received xx xx, 20xx; revised xx xx, 20xx.}
}



\maketitle

\begin{abstract}
Optical communication systems, operating in C--band, are reaching their theoretically achievable capacity limits. An attractive and economically viable solution to satisfy the future data rate demands is to employ the transmission across the full low--loss spectrum encompassing O, E, S, C and L band of the single mode fibers (SMF). Utilizing all five bands offers a bandwidth of up to $\sim$53.5~THz (365~nm) with loss below 0.4~dB/km. A key component in realizing multi--band optical communication systems is the optical amplifier. Apart from having an ultra--wide gain profile, the ability of providing arbitrary gain profiles, in a controlled way, will become an essential feature. The latter will allow for signal power spectrum shaping which has a broad range of applications such as the maximization of the achievable information rate $\times$ distance product, the elimination of static and lossy gain flattening filters (GFF) enabling a power efficient system design, and the gain equalization of optical frequency combs. In this paper, we experimentally demonstrate a multi--band (S+C+L) programmable gain optical amplifier using only Raman effects and machine learning. The amplifier achieves $>$1000 programmable gain profiles within the range from 3.5 to 30~dB, in an ultra--fast way and a very low maximum error of $1.6 \cdot 10^{-2}$~dB/THz over an ultra--wide bandwidth of 17.6--THz (140.7--nm).
\end{abstract}


\IEEEpeerreviewmaketitle

\section{Introduction}
\label{sec:intro}
\IEEEPARstart{O}{ver} the past two decades, a great evolution of optical communication systems, in terms of spectral efficiency$\times$distance product, has been enabled by the advances in digital coherent detection. So far, most of the efforts, on reaching the capacity of the nonlinear fiber--optic channel, have been focusing on the C--band only~\cite{agrell2016roadmap}. However, squeezing the information inside this transmission window will soon reach its theoretical limit~\cite{Dar14}. To cope with the constant demand for higher throughput, novel solutions must be explored.

Optical communication systems operating across multi--band transmission, are an attractive solution for providing the future capacity scaling. They can provide up to 10$\times$ higher capacity, compared to the C--band~\cite{Ferrari20}, on the already deployed SMF fiber infrastructure. To make multi--band systems commercially deployable in the near future, large research efforts in terms of components, system and network design are needed~\cite{ciena, infinera, Doerr2016, Timurdogan19, Messner20, Tummidi20, Sarwar20, Wang20}. 

One of the main challenges in realizing multi--band systems is the development of optical amplifiers that are able to provide sufficiently high gains over such a wide bandwidth. Additionally, a novel feature that may become essential is the ability to provide arbitrary gain profiles in a controlled and ultra--fast way. This is because different signal channels in a multi--band system are unevenly impacted by the interaction between the Kerr nonlinearity, amplified spontaneous emission (ASE) noise and stimulated Raman scattering (SRS)~\cite{Ferrari20}. Consequently, for the maximization of the achievable information rate (AIR) $\times$ distance product, non--flat signal channel power profiles are needed. Depending on the system configuration, signal channel power profiles will be a result of a complex optimization and may assume arbitrary shapes. Moreover, to address the future requirements on high capacity optical networks, ultra--fast gain profile re--configurability is needed~\cite{Napoli18}.

A current and by far the most dominant approach for performing programmable signal channel power profile shaping is by leveraging the use of wavelength selective switches (WSSs) whose primary function is to route the signals throughout the optical network. However, this approach is highly power inefficient since it adjusts the channel powers by means of attenuation.

A novel approach for realizing signal channel power shaping is by employing optical amplifiers with programmable (arbitrary) gain profiles. What we mean by programmable is that the targeted gain profiles can be obtained in a single--step by applying the appropriate pump laser driving voltages. To express it differently, a programmable optical amplifier is an amplifier that can provide arbitrary gain profiles, in a controlled way, with a single--set of instructions. This is somehow equivalent to field-programmable-gate-arrays (FPGAs) in electronics. 

Programmable gain amplifiers could be a potential game changer as they would be able to simultaneously amplify the optical data signal and perform gain shaping. This has many impactful applications such as compensation of wavelength--dependent loss in devices such as modulators and frequency combs, gain--shaping in fixed-gain profile amplifiers and channel power profile adjustments to optimize the AIR in multi--band systems. Especially, if integrated--combs are targeted for multi-channel sources, an efficient approach for gain shaping would be desirable. This is because for integrated--combs there is a large variation in power of their frequency components. Finally, optical amplifiers providing arbitrary gain profiles can be used in hybrid approaches to complement the gain, and overcome the limitations of other optical amplifier technologies~\cite{Fukuchi01,Gordienko16,Ionescu19,Galdino19,Arnould20,Ye20}. 

There are several approaches and technologies for realizing optical amplifiers covering multiple bands. 
To date, works on multi--band optical amplifiers have focused on: rare--earth--doped fiber amplifiers (xDFAs) covering 17.56~THz over O+E--band~\cite{Wang20} and 10.7~THz over S+C--band~\cite{Sakamoto06}, semiconductor optical amplifiers (SOAs) for 12.7~THz on S+C+L--band~\cite{Renaudier18}, optic parametric amplifiers (OPAs) with 10~THz of bandwidth on S+C+L--band~\cite{Kobayashi20}, Raman amplifiers (RA) in combination with EDFAs, SOAs and OPAs achieving bandwidths ranging from 10.7 to 14~THz on C+L and S+C+L--band~\cite{Fukuchi01,Ye20,Ionescu19,Galdino19,Arnould20,Gordienko16}, and pure RAs with bandwidths of up to 19.1--THz S+C+L--band~\cite{Rottwitt99,Zhou06,Chen18,Emori01,Iqbal20}. So far, the majority of works in~\cite{Fukuchi01,Sakamoto06,Wang20,Renaudier18,Kobayashi20,Gordienko16,Ionescu19,Galdino19,Arnould20,Rottwitt99,Zhou06,Chen18,Emori01,Iqbal20} have focused on realizing flat gain profiles in C+L and S+C+L--band. Recently, an amplifier that relies on a hybrid SOA/Raman configuration has been demonstrated to achieve arbitrary loss/gain profile generation in S+C+L--band in 12.3~THz of bandwidth~\cite{Ye20}.

Among all different solutions, RAs are most suitable for realizing arbitrary gain profiles, in a controlled way. This is because the RAs allow for a flexible gain profile design by adjusting the pump powers and wavelengths, and provide gain availability across a broad range of wavelengths, when operated in multi-pump configurations.

The challenge with Raman amplifier design is on the selection of pump powers and wavelengths that would result in a targeted gain profile. Several solutions to this optimization problem have been reported in the literature but have mainly focused on realizing flat gain profiles~\cite{Ferreira11,Zhou01,Chen18,Perlin02,Iqbal20,Mowla08,Jiang10,Emori01, Ania07}. Recently, a machine learning framework for the ultra--fast configuration of the pump powers and wavelengths has been theoretically proposed and as a proof--of--principle experimentally demonstrated in C--band only~\cite{Zibar20,deMoura20}. The proposed approach can be used for the design of Raman amplifiers, where an arbitrary gain profile is achievable in a controlled way. However, moving from C--band to multi--band and realizing wider gain profiles is significantly more challenging. This is partly due to the increased number of pumps that need to be controlled and also the increased nonlinearity given the higher overall powers in the optical fiber.

In this paper, we use the proposed machine learning framework for the experimental realization of multi--band RAs that can provide arbitrary gains, in a controlled way, in C+L and S+C+L--band. Up to 8 pumps are employed to provide more than 5000 arbitrary gain profiles over up to 17.6-THz of bandwidth. We achieve a highly--accurate programmable set of gain profiles with a very low average maximum error, (defined between the target and realized gain profiles), per bandwidth, $E_{MAX}/BW$, of $1.6 \cdot 10^{-2}$~dB/THz.

This is the first experimental demonstration of S+C+L--band optical amplifier, that can realize arbitrary gain profiles in a controlled way, using Raman effects only. We have achieved an important breakthrough by demonstrating an extremely low root mean square error per bandwidth (RMSE/BW) of 0.0045 dB/THz, over an ultra--wide bandwidth of 17.6 THz (140.7 nm). More specifically, in terms of maximum error per bandwidth, our results are a record low. The presented approach and the obtained results have therefore great potential to become a relevant reference point for future research on this upcoming topic.

The previous experimental results that we have published in \cite{Zibar20,deMoura20} were limited to the C--band only. Increasing the bandwidth from C to S+C+L--band (a factor of 4.4 for the considered case) is highly--challenging. We demonstrate that the proposed machine learning framework plays a key role in addressing those challenges.

Machine learning for broadband gain optimisation is a topic of growing interest, which is reflected in the recent work~\cite{Ye20} reporting a root mean squared error per bandwidth, $RMSE/BW$, of 0.033~dB/THz in a 12.3~THz bandwidth SOA/distributed Raman link scenario. This is an order of magnitude higher $RMSE/BW$ when compared to our current result of 0.0045 dB/THz over a larger bandwidth of 17.6~THz. 

\label{sec:exp_setups}
\begin{figure*}[t]
  \centering
  \includegraphics[width=1\textwidth]{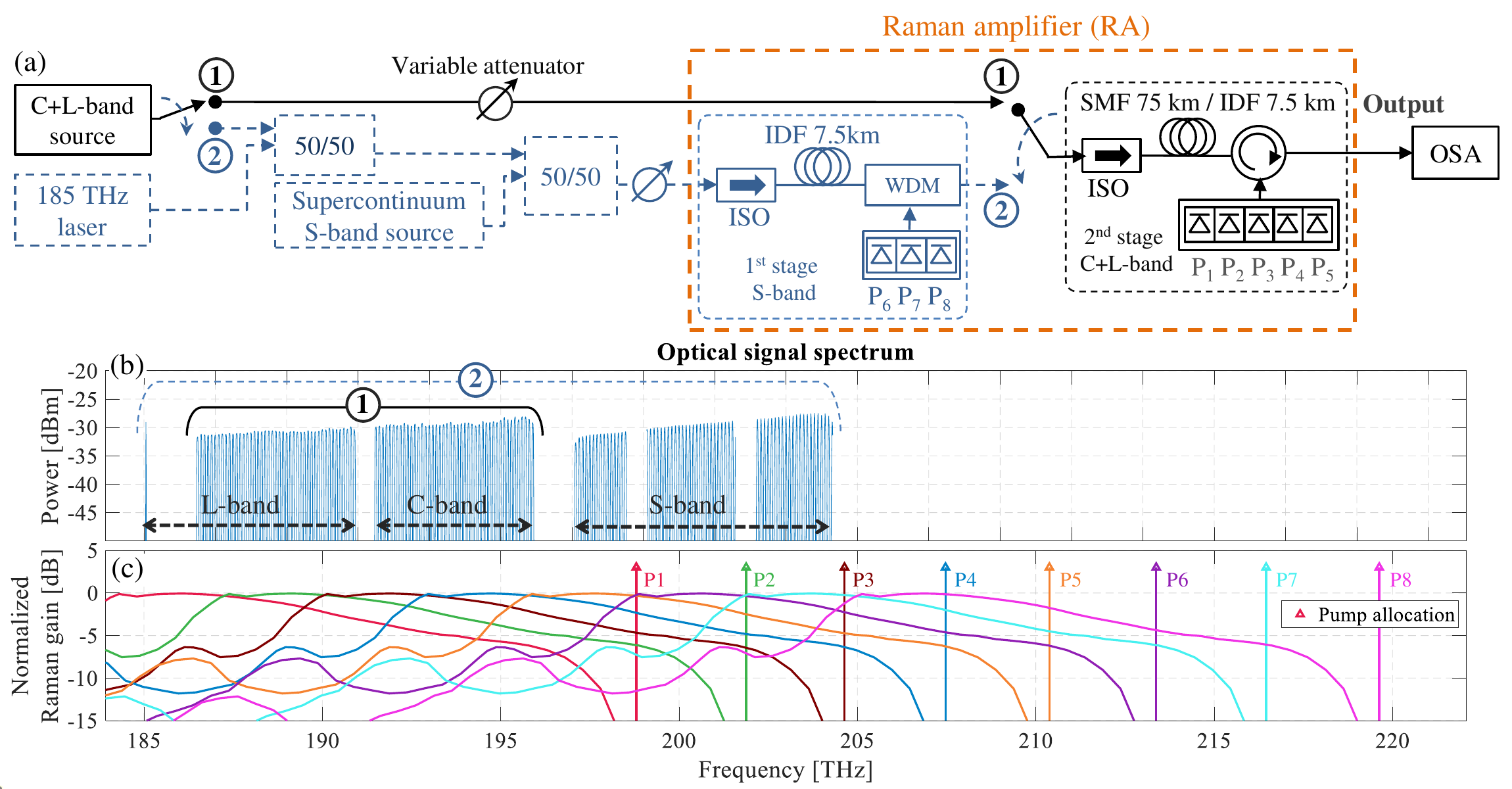}
\caption{(a) Experimental setup for the multi--band RA: path 1 refers to the C+L--band RA and path 2 is for the S+C+L--band dual--stage discrete RA. (b) Input optical signal spectrum. (c) Pump lasers spectrum and their expected contribution to the overall Raman gain.}
\label{fig:setup}
\end{figure*}

The structure of the paper is as follows: Section~\ref{sec:exp_setups} describes the experimental setup for realizing Raman amplifiers operating in C+L and S+C+L--band. We also give a brief overview of the ML framework used to obtain programmable arbitrary gain profiles. Section~\ref{sec:results} presents, discusses and evaluates the experimental results. In Section~\ref{sec:conc} conclusions and future work are presented and outlined.

\section{Experimental setup}
The experimental setup for realizing the multi--band RA is shown in Fig.~\ref{fig:setup}(a). By selecting path 1 or 2, the operation in either C+L (1) or S+C+L--band (2) can be enabled. To achieve gains in the C+L and S+C+L--band, 5 and 8 pump lasers are employed, respectively. Fig.~\ref{fig:setup}(c) illustrates the spectral pump allocation and their individual contribution to the overall Raman gain. We only consider counter propagating pumps whose wavelengths are fixed and shown in Table~\ref{tab:pumps}.
 
\begin{table}[!b]
\centering
\caption{\bf Pump lasers wavelengths and frequencies}
\begin{tabular}{ccccc}
\hline
 & $P_1$ & $P_2$ & $P_3$ & $P_4$ \\
\hline
Wavelength [nm] & 1508 & 1485 & 1465 & 1445 \\
Frequency [THz] & 198.8 & 201.9 & 204.6 & 207.5 \\
\hline
 & $P_5$ & $P_6$ & $P_7$ & $P_8$ \\
\hline
Wavelength [nm] & 1425 & 1405 & 1385 & 1365 \\
Frequency [THz] & 210.4 & 213.4 & 216.5 & 221.1 \\
\hline
\end{tabular}
\vspace{1ex}
  \label{tab:pumps}
\end{table}

The gain profile control is performed by only adjusting the pump powers. Pump lasers $P_1...P_7$ are semiconductor laser diodes. Their output power is controlled by adjusting the driving currents. The corresponding power going into the RA is in the range from $\sim$16~dBm to $\sim$27~dBm. Pump laser $P_8$ is a Raman--based fiber laser and is controlled by adjusting its voltage. It provides power to the RA ranging from $\sim$20~dBm to $\sim$27~dBm.

The reason why we only optimize pump lasers powers is because there are no tunable, high power pump lasers available within the considered frequency ranges. However, the selected pump laser frequencies fall within the ranges that would provide Raman gain profiles within the desired frequency bands.

\subsection{C+L--band Raman amplifier}
The C+L--band RA can either be operated as a discrete, (7.5~km of inverse dispersion fiber (IDF)) or distributed (75~km span of standard SMF) amplifier. An input optical signal covering the C+L--band, for testing the performance of the RA, is generated by using two ASE sources for C and L bands channelized through a WSS to generate 90 lines placed at 100~GHz ITU-T grid covering a 9.4~THz (77~nm) bandwidth. The total input signal power to the amplifier is adjusted by means of a variable attenuator to 0 and 10~dBm for the discrete and distributed C+L--band Raman amplifier, respectively. The corresponding optical spectrum is shown in Fig.~\ref{fig:setup}(b) (inside bracket 1) and is measured with a resolution of $\Delta \lambda = 0.1$~nm. The gaps between the C and L signal bands are due to the different ASE sources for these two bands. 
An isolator is placed at the input to the IDF to prevent pump powers entering the C+L--band signal source and to minimize the double Rayleigh backscattering induced multipath interference~\cite{Iqbal19JLT}. Finally, an optical spectrum analyser (OSA) is used to capture the optical spectrum.

\subsection{S+C+L--band Raman amplifier}
The S+C+L--band RA is implemented as a two--stage sequential discrete RA. The first stage is responsible for providing the gain in the S--band and it consists of 7.5~km of IDF and three pump lasers, $P_6...P_8$ used to control the gain profiles. The second stage is the same as the one used for the C+L--band RA. Note that distributing the pumps into two sequential stages reduces the strong depletion of shorter wavelength pumps~\cite{Krummrich01}. The multi--band input optical signal (17.6~THz/140.7~nm) is generated by combining the optical signal from the C+L--band with a supercontinuum S--band source~\cite{El-Taher:s} and a single frequency laser operating at 185~THz. The resulting signal has a total of 148 frequency lines at 100~GHz ITU-T grid. A variable attenuator is used to adjust the input signal power to 7~dBm. The corresponding optical spectrum is shown in Fig.~\ref{fig:setup}(b) (inside bracket 2). Due to the amplifier configuration, two pumps from the first stage ($P_{1-2}$) fall within the S-band signal. This means that some channels from the S--band need to be removed to avoid overlapping with the Rayleigh backscattered components of the pumps, leaving the gaps as shown in Fig.~\ref{fig:setup}(b)~\cite{Iqbal:20}.

\subsection{Pump power control}
\begin{figure*}[t]
  \centering
  \includegraphics[width=0.32\textwidth]{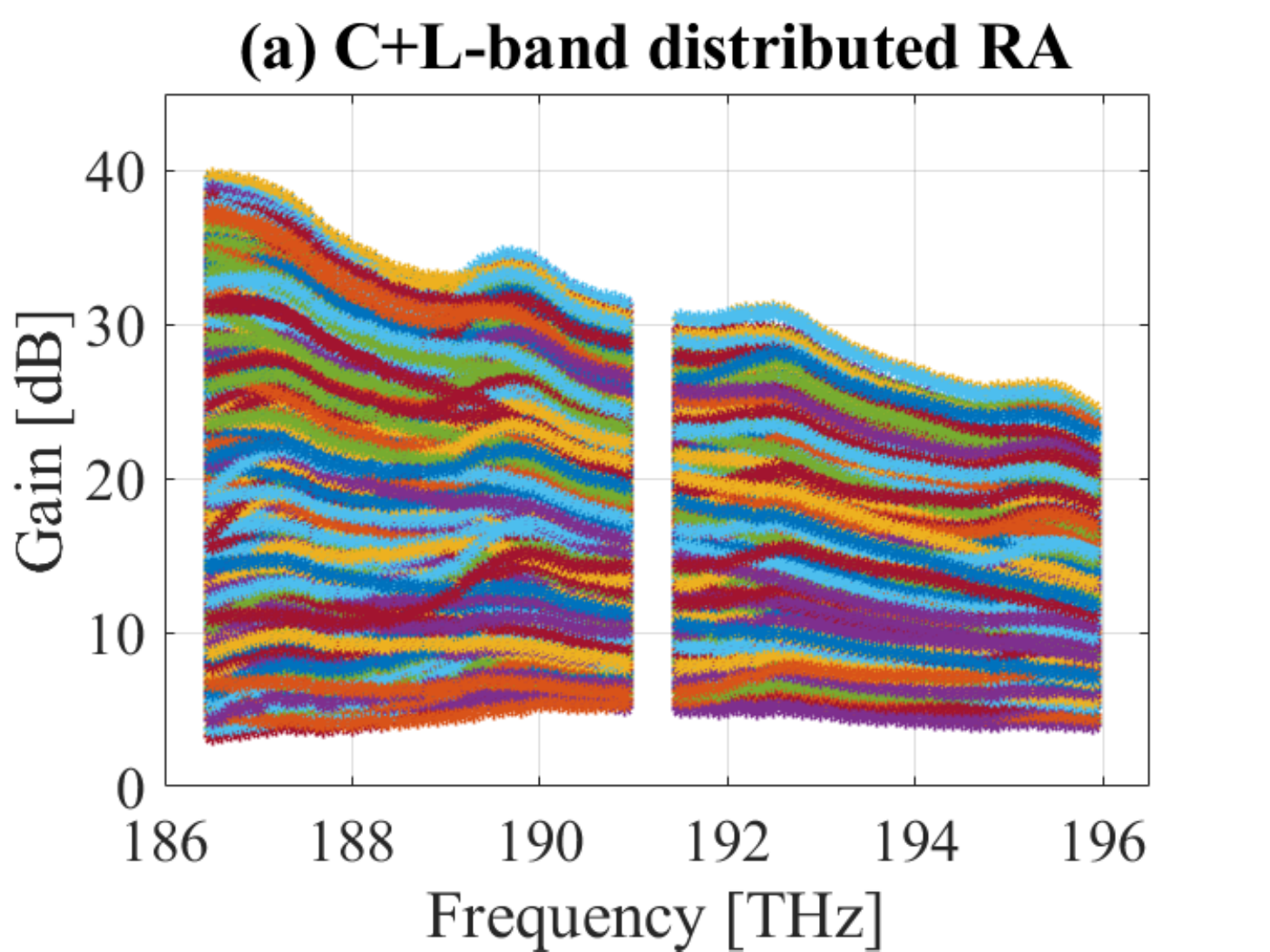}
  \includegraphics[width=0.32\textwidth]{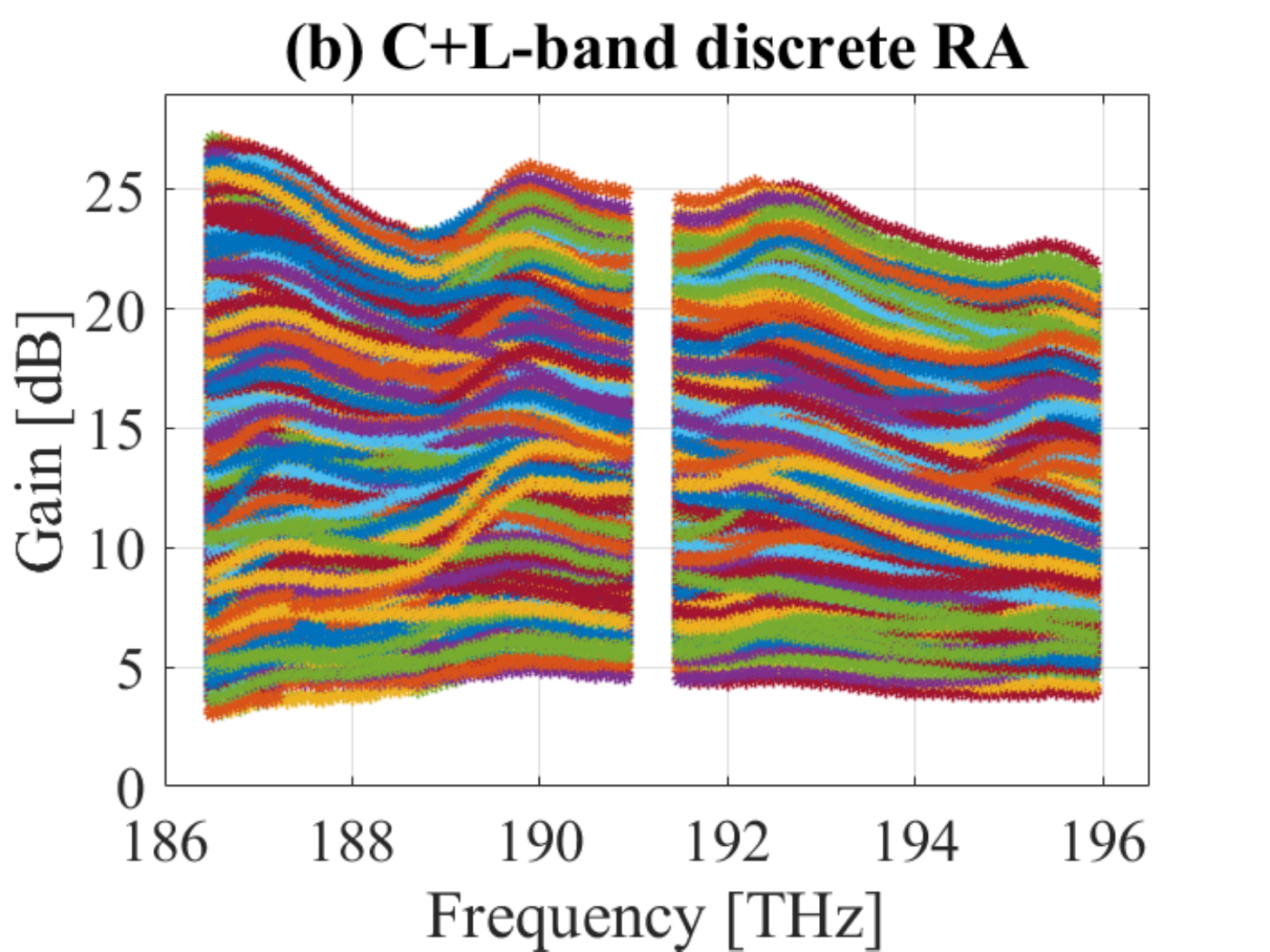}
  \includegraphics[width=0.32\textwidth]{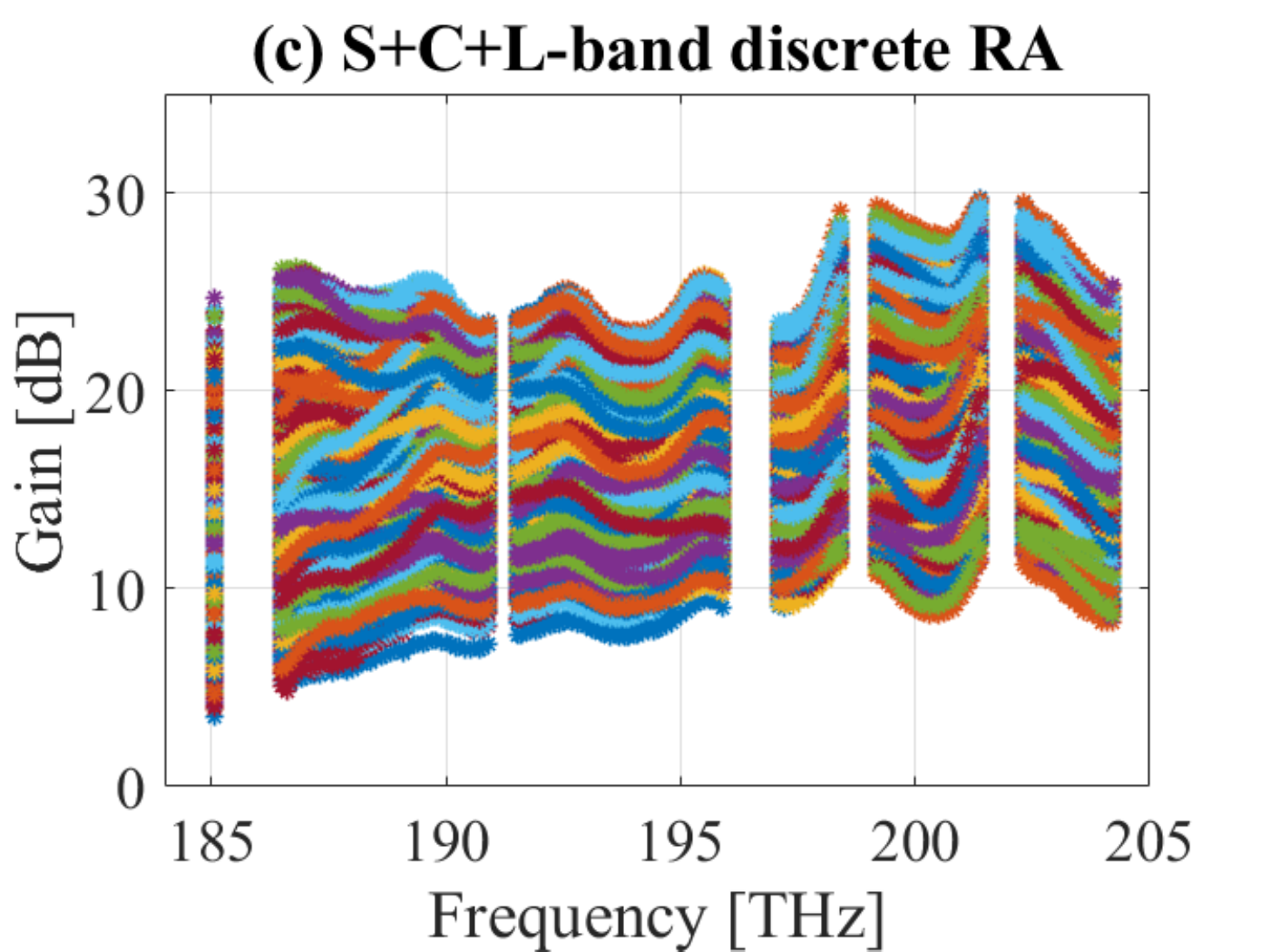}
\caption{Measured on--off gain profiles for various pump laser currents and voltage configurations. C+L--band RA (a) distributed, (b) discrete and (c) dual stage discrete RA S+C+L--band.}
\label{fig:datasets}
\end{figure*}

The objective is to determine pump power settings that result in user defined target gain profiles such as: tilted gain, flat gain or an arbitrary gain. These settings are achieved off--line using the machine learning framework presented and then later applied on--line for the pump laser currents and voltage control~\cite{Zibar20}. As the framework in~\cite{Zibar20} is based on supervised learning, a data--set is required. This is achieved by varying the currents and the voltage of the pump lasers and measuring the corresponding gain profiles. The gain profiles are measured on a 100-GHz grid, as the difference in power between the output optical spectrum when the pump lasers are turned on and off, also known as the on--off gain. As the currents and the voltage, $I_1,...I_7,V_8$, are drawn from a uniform distribution whose bounds are shown in Table~\ref{tab:pump_range}, we refer to the corresponding gain profiles as arbitrary. 

In Fig.~\ref{fig:datasets}, the measured on--off gain profiles, $G$, obtained for the C+L and the S+C+L--band are shown. For the discrete Raman amplifiers, increasing the pump laser output powers, beyond certain levels, leads to gain instabilities. The maximum allowable driving currents, for the pump lasers, are shown in Table~\ref{tab:pump_range} as the maximum values of the uniform distribution interval. As a consequence of the limited pump lasers output powers on the discrete C+L--band amplifier, we could not obtain as large gains and gain profile variations as compared to the distributed C+L--band amplifier. More specifically, the decreased driving currents for the lower frequency pumps ($P_1$ and $P_2$) is the reason why the gains in lower frequency region in Fig.~\ref{fig:datasets}(b), (186 THz--188 THz), are not as high as for the distributed amplifier (Fig.~\ref{fig:datasets}(a)). Additionally, the reduced current on the low frequency pump leads to a lower depletion experienced by the high frequency pumps.

We measure $M=5600$ and $M=4025$ gain profiles, each with $K=90$ and $K=148$ data points per gain profile, for C+L and S+C+L--band, respectively. We denote the respective data--sets as: $\mathcal{D}^{M\times (K+5)}_{C+L}=\{(G_1^i,...,G_K^i,I^i_1,...I^i_5),|i=1,...,M\}$ and $\mathcal{D}^{M\times (K+8)}_{S+C+L}=\{(G_1^i,...,G_K^i,I^i_1,...I^i_7,V^i_8),|i=1,...,M\}$.

\begin{table}[!h]
\centering
\caption{\bf {Current and voltage ranges}}
\begin{tabular}{cccc}
\hline
            & C+L dist. & C+L disc. & S+C+L disc.           			\\
\hline
$I_1$ [A]  & $[0.20:1.00]$   	& $[0.20:0.90]$  	& $[0.20:1.00]$   	\\
$I_2$ [A]  & $[0.20:1.00]$   	& $[0.20:0.80]$  	& $[0.20:0.80]$    	\\
$I_3$ [A]  & $[0.20:1.20]$   	& $[0.20:1.20]$ 		& $[0.20:1.00]$   	\\
$I_4$ [A]  & $[0.20:1.50]$   	& $[0.20:1.40]$ 		& $[0.20:1.40]$   	\\
$I_5$ [A]  & $[0.20:1.50]$   	& $[0.20:1.50]$ 		& $[0.20:1.40]$   	\\
$I_6$ [A]  & - 				& - 					& $[0.20:1.20]$ 	\\
$I_7$ [A]  & - 				& - 					& $[0.60:1.30]$		\\
$V_8$ [V]  & - 				& - 					& $[1.80:2.40]$ 	\\
\hline
\end{tabular}
\vspace{1ex}
  \label{tab:pump_range}
\end{table}

To find the machine learning model with the lowest prediction error, we allocate 3400 and 3000 data points, for C+L and S+C+L--band, correspondingly. We employ 10--fold cross--validation, which means that we use 90\% for training (includes hyperparameter optimization) and 10\% for testing as described~\cite{Zibar20,bishop2006}. For a more detailed explanation on the training of the employed machine learning model, see the Appendix Section. The remaining data points are later used for the final validation of the machine learning model for the pump laser current prediction of arbitrary gains.

The procedure of obtaining pump current configuration is then as follows: 1) a single-layer neural network, $NN_{inv}$, is employed to learn the mapping between the target gain profiles and currents and voltage -- inverse system learning, 2) once the neural network has learned the inverse mapping, given a set of target gain profiles, the corresponding pumps currents and voltages are predicted, 3) the predicted currents and voltages are then applied to the second multi--layer neural network, $NN_{fwd}$, that has learned the forward mapping between pump currents/voltage and gain profiles. The $NN_{fwd}$ thereby predicts the gain profile given the pump currents and the voltage. If the error between the predicted and targeted gain profile is not satisfactory pump currents and voltages are adjusted accordingly, i.e.~fine--optimization. The fine--optimization uses iterative gradient descent by backpropagating the error through $NN_{fwd}$ to adjust the currents and voltage as described in~\cite{Zibar20}, 4) the obtained currents and voltages are applied to the pump lasers in the experimental set--up, and new sets of measurements are performed, and 5) finally, to investigate the accuracy of the predicted pump currents and voltage, we calculate the maximum absolute error between the target and the newly measured gain profiles (i.e.~$E_{MAX}$) and normalize it with the bandwidth ($BW$). The optimized topologies of the employed neural networks $NN_{fwd}$ and $NN_{inv}$, as well as their performance evaluation, are found in the Appendix Section.

\section{Results and discussion}
\label{sec:results}
\subsection{Arbitrary gain profiles}
\begin{figure*}[!t]
  \centering
  \includegraphics[width=0.32\textwidth]{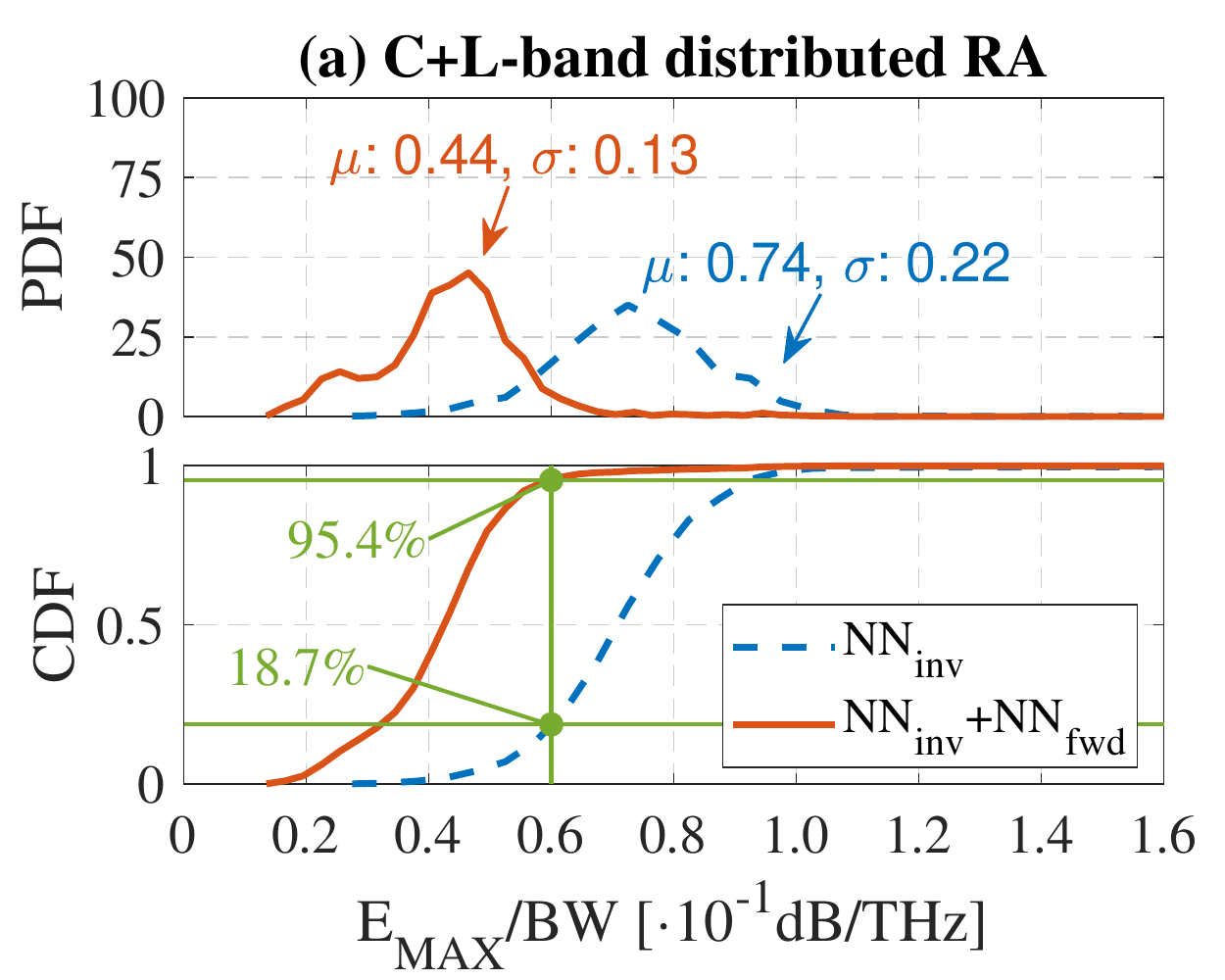}
  \includegraphics[width=0.32\textwidth]{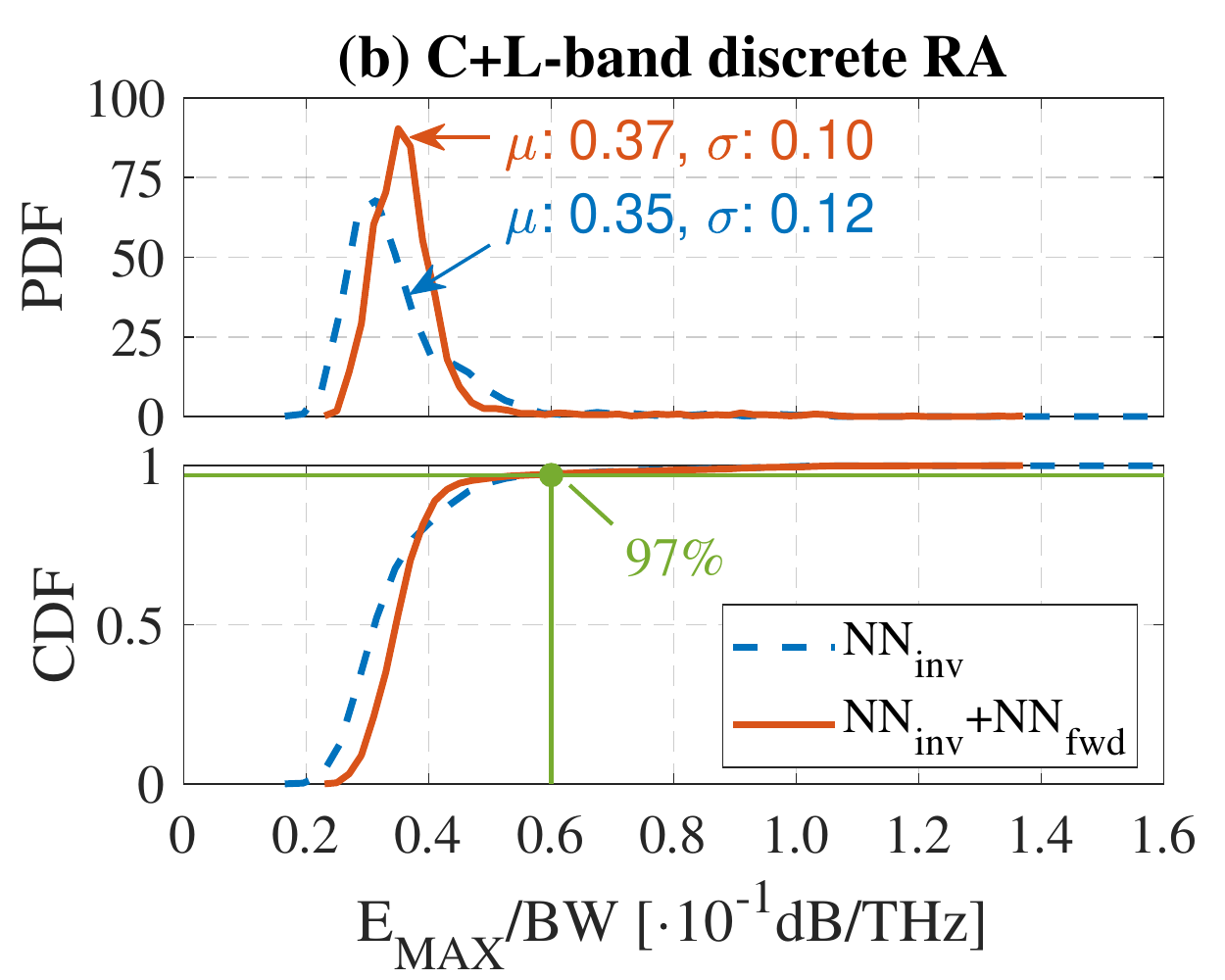}
  \includegraphics[width=0.32\textwidth]{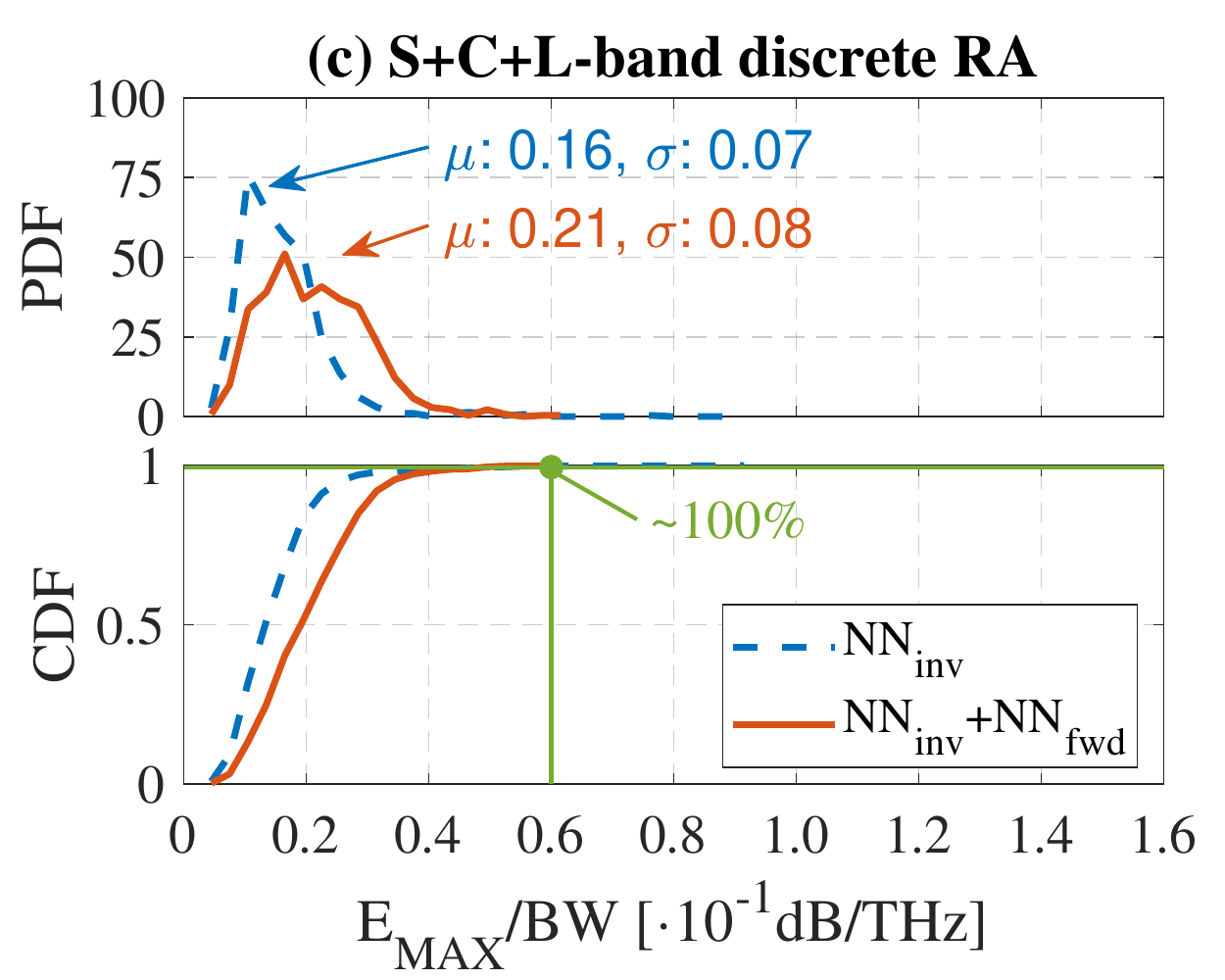}
\caption{Probability density function (PDF, top) and cumulative density function (CDF, bottom) of the $E_{MAX}/BW$, with indication of mean, $\mu$ and standard deviation, $\sigma$: (a) C+L--band distributed RA, (b) C+L--band discrete RA and (c) S+C+L--band discrete RA.}
\label{fig:results_arb}
\end{figure*}

Fig.~\ref{fig:results_arb}(a)--(c) show the probability, (PDF), and the cumulative, (CDF), density functions of the $E_{MAX}/BW$ for the C+L--band (distributed and discrete) and S+C+L--band (discrete) Raman amplifiers. The error is defined between the targeted arbitrary gain profiles, taken directly from the data--set (not used for training the machine learning framework), and the predicted gain profiles obtained from the measurement using the pump currents and voltage allocation provided by the machine learning framework. We use 2100, 2600 and 1025 target arbitrary gain profiles for the distributed C+L--band, discrete C+L--band and discrete S+C+L--band validation, respectively. We compare the accuracy of allocating pump currents and voltage, by using only the inverse mapping multi--layer neural network, $(NN_{inv})$, and both the inverse and forward mapping multi--layer neural networks, $(NN_{inv}+NN_{fwd})$, which allows for fine--optimization of pump currents and the voltage.

\begin{figure*}[!t]
  \centering
  \includegraphics[width=0.32\textwidth]{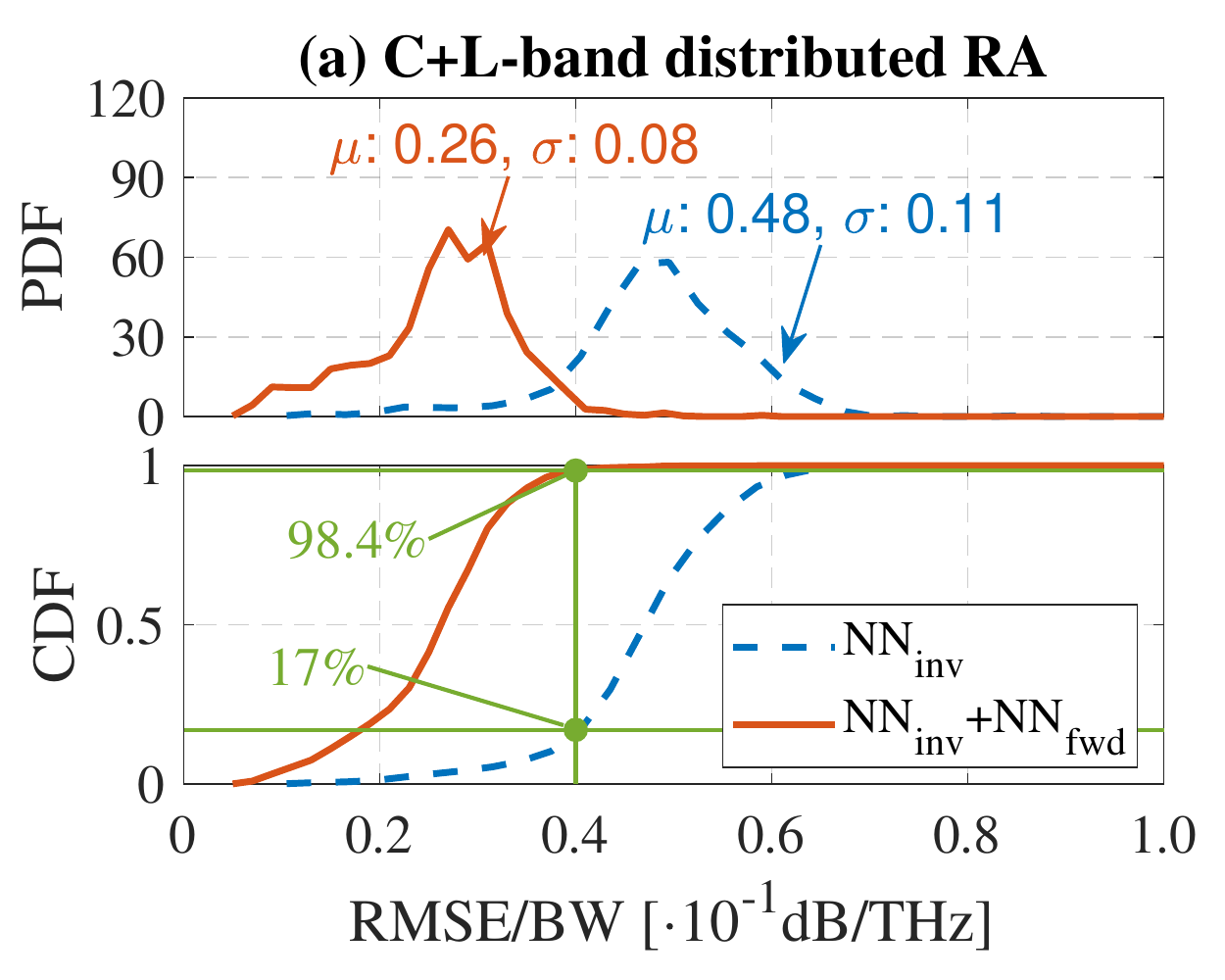}
  \includegraphics[width=0.32\textwidth]{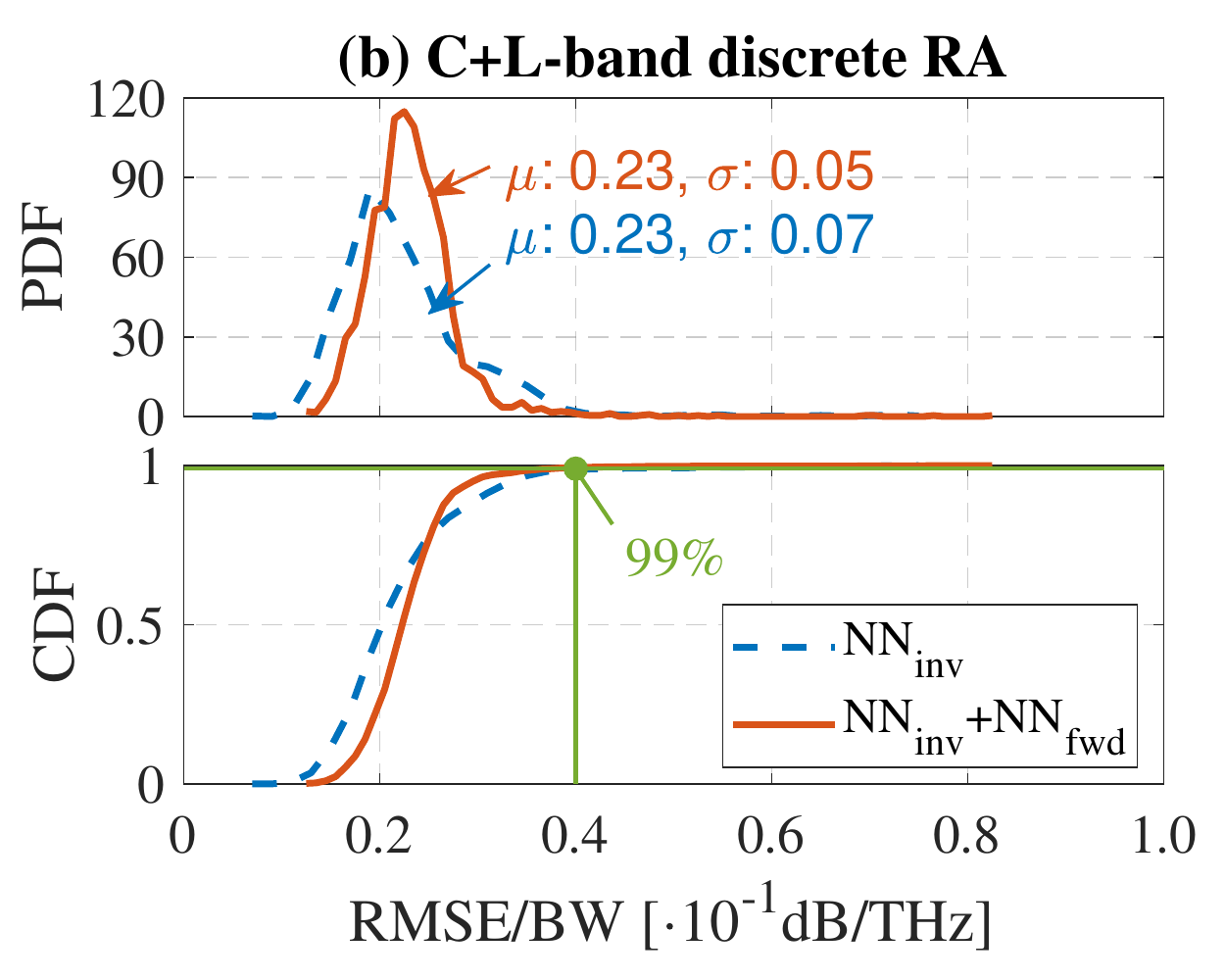}
  \includegraphics[width=0.32\textwidth]{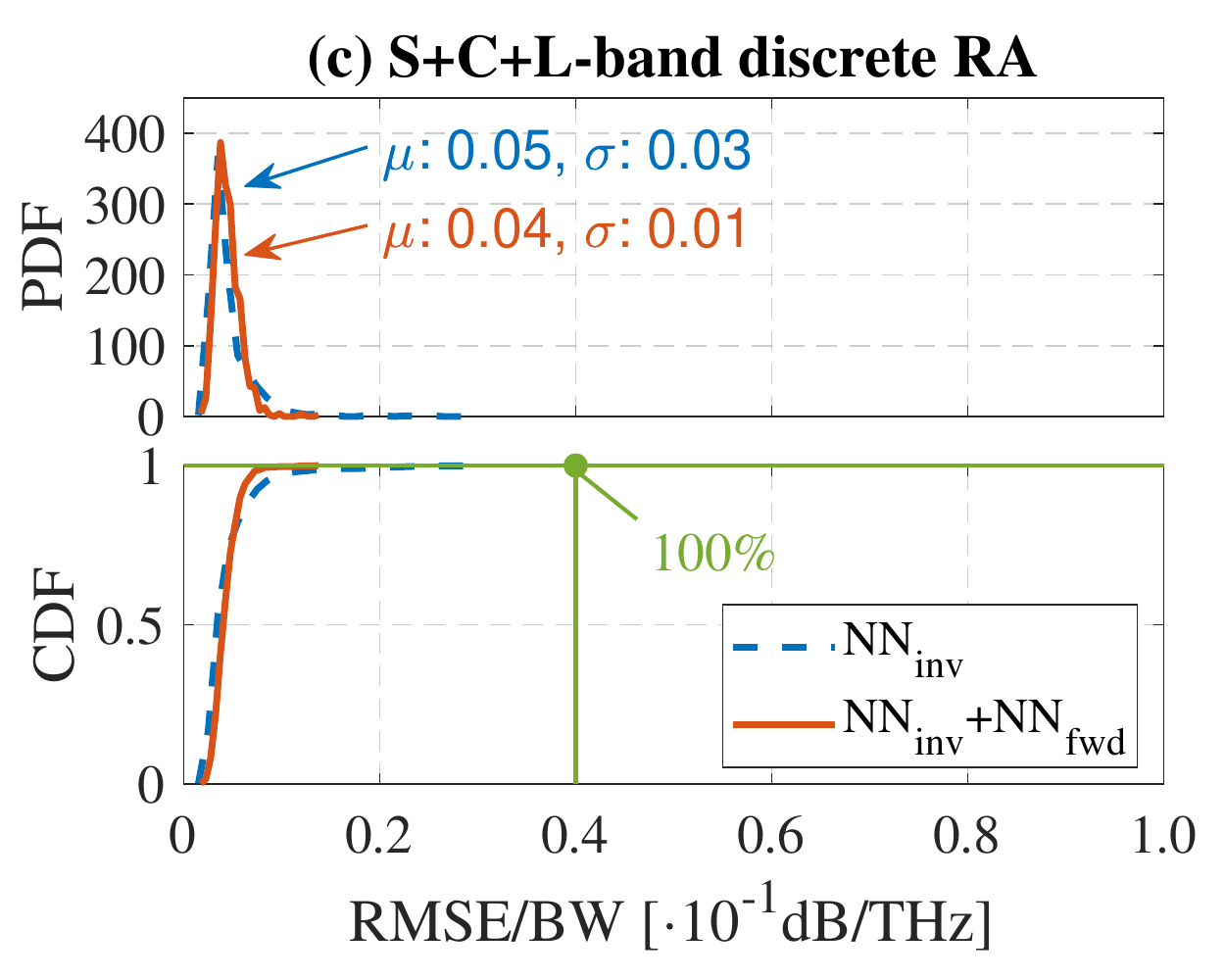}
\caption{Probability density function (PDF, top) and cumulative density function (CDF, bottom) of the $RMSE/BW$, with indication of mean, $\mu$, standard deviation, $\sigma$: (a) C+L--band distributed RA, (b) C+L--band discrete RA and (c) S+C+L--band discrete RA.}
\label{fig:results_arb_rmse}
\end{figure*}

The PDFs shown in Fig.~\ref{fig:results_arb}(b)--(c), illustrate that for the discrete RA, highly--accurate pump current predictions, resulting in a low mean and standard deviation, can be obtained using only $NN_{inv}$. Thus, the currents and the voltage prediction is obtained in an ultra--fast way as $NN_{inv}$ only involves matrix computations. We notice that the mean and standard deviations are decreased by a factor of $\sim$2 when going from C+L to S+C+L--band. This is mainly because these two schemes have the same performance in terms of $E_{MAX}$ and S+C+L--band has almost two times wider bandwidth. However, qualitatively the results for C+L and S+C+L--band are comparable.   

If $NN_{inv}+NN_{fwd}$ is used a slight increase in the mean and the standard deviation is observed. This is because the $NN_{inv}$ has already found pump current configuration that minimizes the mean square error. Applying the fine--optimization introduces some small random deviations around this minimum and worsens the performance. 

For both discrete RA schemes, the CDF shows that most of the cases already present an $E_{MAX}/BW$ lower than $6\cdot10^{-2}$~dB/THz, before the fine--optimization, i.e.~97\% of the cases for the C+L--band and $\sim$100\% for the S+C+L--band.  

Compared to the discrete RA, the resulting PDF for the distributed RA (Fig.~\ref{fig:results_arb}(a)) has a higher mean and standard deviation when considering only $NN_{inv}$. On the other hand, a significant reduction can be obtained after applying fine--optimization $NN_{inv}+NN_{fwd}$, as also illustrated by the CDF. Indeed, the fine--optimization significantly increases the number of cases with $E_{MAX}/BW$ lower than $6\cdot10^{-2}$~dB/THz, i.e.~from 18.7\% to 95.4\%.

To understand why only the distributed amplifier benefits from the fine--optimization, we need to consider the mean and the standard deviation of the predicted RMSE for the arbitrary gain profiles when applying $NN_{inv}$ only. This information is obtained from Fig.~\ref{fig:results_arb_rmse}(a)--(c) by de-normalizing it with the amplifier bandwidth. The corresponding mean and standard deviations, ($\mu \pm \sigma$), for the distributed C+L--band, discrete C+L--band and discrete S+C+L--band amplifier are: $0.46 \pm 0.10$ dB, $0.21 \pm 0.06$ dB and $0.08 \pm 0.05$ dB. As the RMSE values for the discrete C+L-band and, especially, discrete S+C+L--band amplifier are already low, there are no observable improvements when applying the fine--optimization.

Finally, in Fig.~\ref{fig:results_arb_rmse}(a)-(c), the resulting PDF and CDF of the RMSE per bandwidth is plotted for the distributed and discrete amplifiers. The Figure shows that very low mean and standard deviation values are achievable.

\subsection{Flat and tilted gain profiles}
\begin{figure*}[!t]
  \centering
  \includegraphics[width=0.32\textwidth]{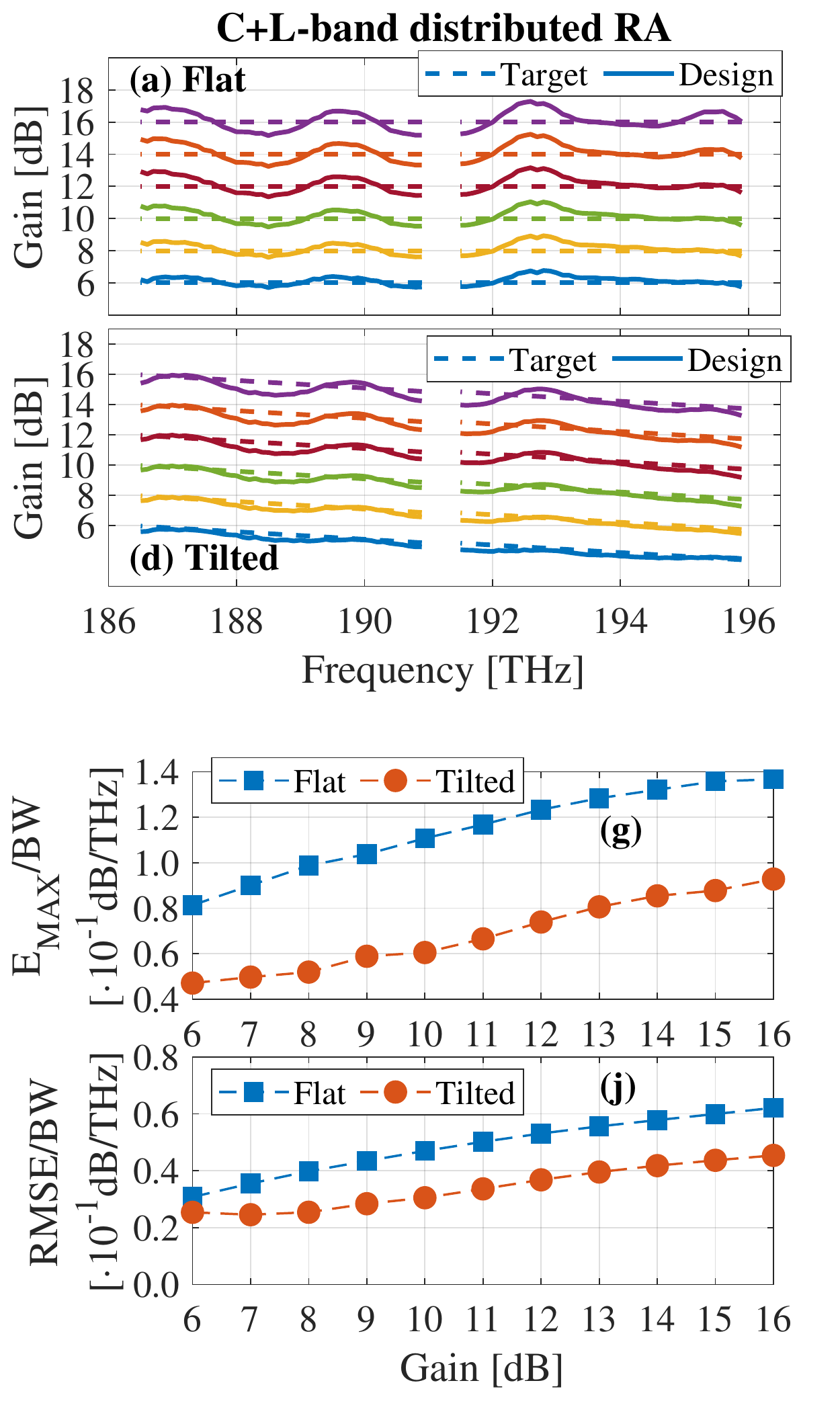} 
  \includegraphics[width=0.32\textwidth]{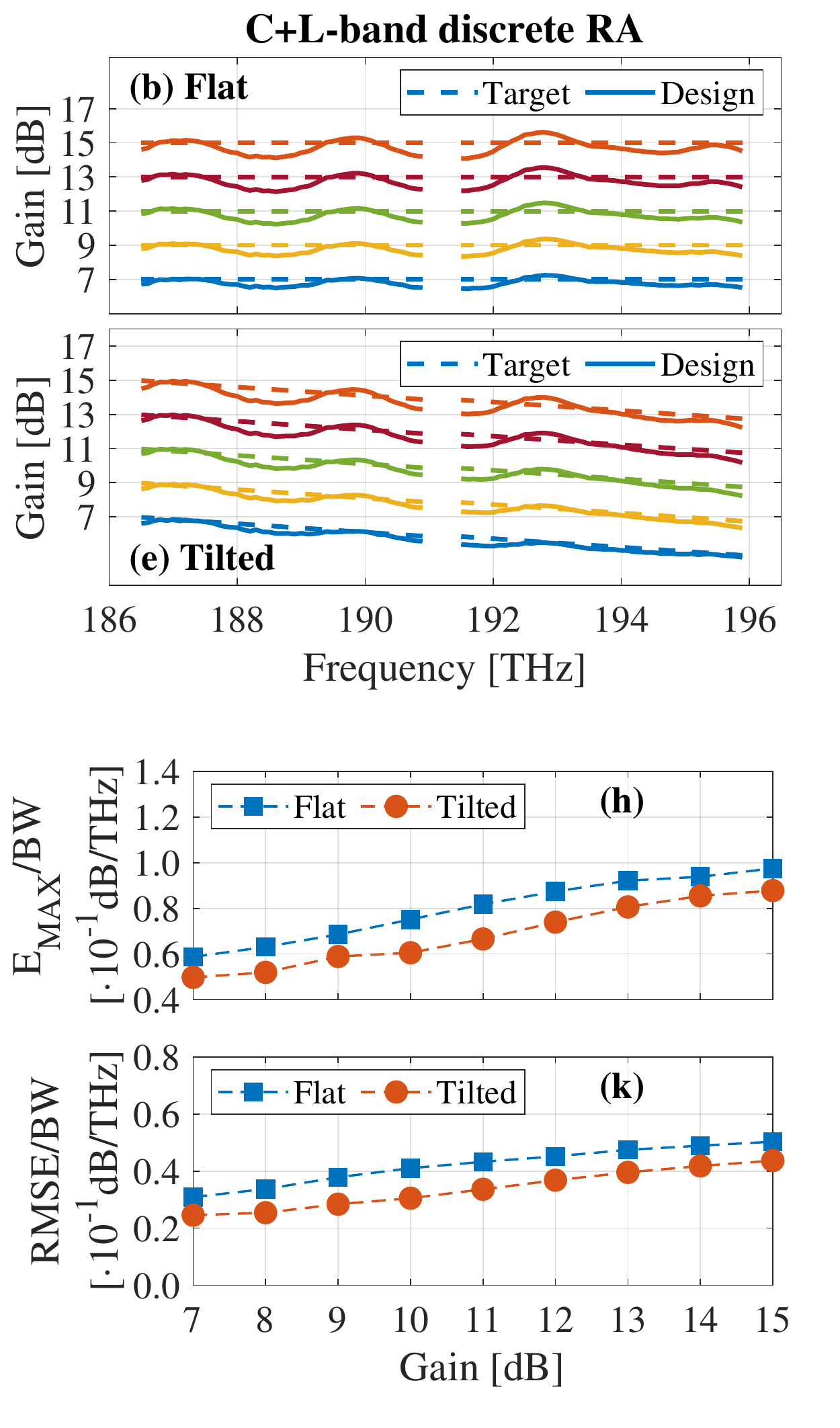}
  \includegraphics[width=0.32\textwidth]{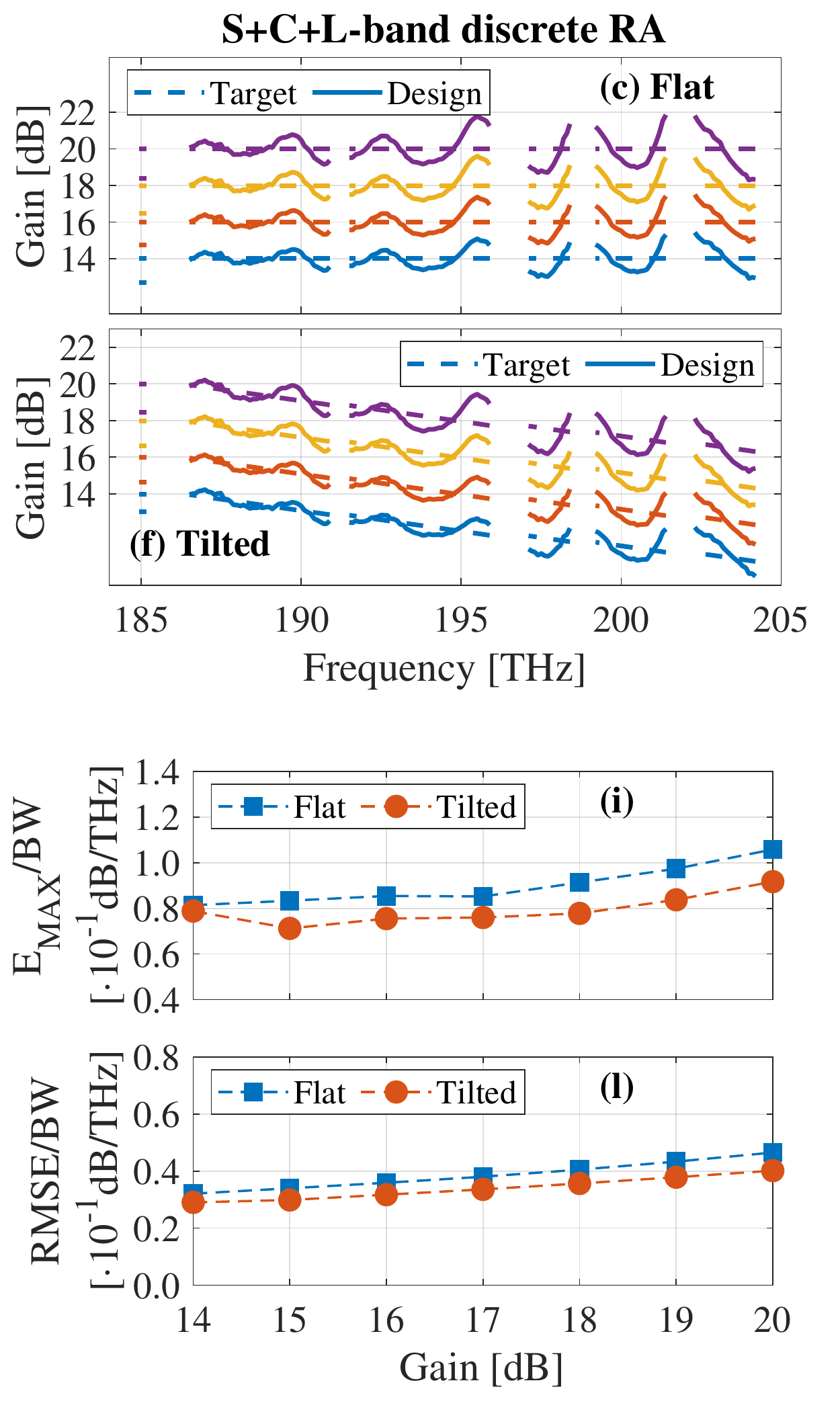}
\caption{(a)--(f): the predicted and the target flat and tilted (on--off) gain profiles as a function of wavelength. (g)--(i) $E_{MAX}/BW$ and (j)--(l) $RMSE/BW$ as a function of gain for the flat and the tilted gains.}
\label{fig:results_flat_tilted}
\end{figure*}
Next, we investigate the ability of the machine learning framework to predict accurate pump current and voltage allocations for the design of flat and tilted gain profiles using the discrete and distributed RAs, in C+L and S+C+L--band. Flat gains ranging from 6 to 16~dB (C+L--band distributed RA), 7 to 15~dB (C+L--band discrete RA), and 14 to 20~dB (S+C+L--band discrete RA) are evaluated in steps of 1~dB. For the tilted profiles, slopes of approximately 0.24~dB/THz (C+L--band RAs) and 0.20~dB/THz (S+C+L--band RA) are considered. These values were chosen to provide an overall tilt of around 1~dB on each band.

Fig.~\ref{fig:results_flat_tilted}, shows the predicted and target flat ((a)-(c)) and tilted ((d)-(f)) gain profiles, as a function of frequency, for the distributed and the discrete RA operating in C+L and S+C+L--band. Just a subset of gains (2~dB step) is shown for better visualization. The corresponding $E_{MAX}/BW$ for all gains under consideration is shown in Fig.~\ref{fig:results_flat_tilted}(g)-(i). We only show results obtained after using $NN_{inv}+NN_{fwd}$ as the fine--optimization significantly reduced the error for all the amplifier schemes and their evaluated gains.

The reason why $NN_{inv}$ is not able to provide accurate solutions for the flat and tilted gain profiles is because, in general, multi--layer neural--networks are good at interpolating and not so good at extrapolating. More precisely, the neural--networks will provide highly accurate predictions for examples that are close to the examples in the training data set. The number of cases in the training data--set with a ``close to flat gain`` profiles ($max([G_1,...,G_K])-min([G_1,...,G_K])\leq$ 1.2 dB), out of the total data--set size, are: 13/3464, 61/3000, 0/3000, for the distributed C+L, discrete C+L and discrete S+C+L--band amplifiers, respectively. These low ``close to flat gain`` profile cases on the training data--set is an indication that the $NN_{inv}$ is extrapolating when predicting the pump configuration for these flat gains. The same analysis goes to the tilted gain profiles.

Additionally, for the increasing input dimension of the neural network, an increasing number of training data points is needed to cover all the combinations. This is explained in details in~\cite{Zibar20,bishop2006}. The input dimension of the $NN_{inv}$ for the S+C+L and C+L--band amplifiers are 148 and 90, respectively (Table~\ref{tab:NNinv} in the Appendix Section). The $NN_{fwd}s$ used on the fine--optimization routine, on the other hand, have a significantly smaller input dimensions compared to the $NN_{inv}s$, i.e.~5 for the distributed and the discrete C+L and 8 for the discrete S+C+L--band amplifier (Table~\ref{tab:NNdir} in the Appendix Section). This implies that they are easier to train and can provide accurate predictions when trained over a smaller data--set size. That is why the fine--optimization is able to outperform $NN_{inv}$ even using another neural network $NN_{fwd}$ trained over the same data--set.

Furthermore, we would like to stress that by employing the fine--optimization, large data sets for training $NN_{inv}$ are not necessary as we are still able to obtain highly--accurate gain designs.

A general trend observed in Fig.~\ref{fig:results_flat_tilted}(a)--(f), is that the predicted gain oscillates around the target gain profile. The magnitude of the oscillations has a tendency to increase for increasing gains. Moreover, for the S+C+L--band RA, the oscillation amplitude increases with the frequency, achieving up to 2~dB of maximum error compared to the target.

To understand what is happening, it is worth mentioning that it was observed some power instabilities on the supercontinuum S--band source and the Raman-based fiber laser $P_8$. Additionally, recall that the broadband and nonuniform Raman gain spectrum for a single pump, with a peak located near 12.5~THz below the pump frequency for the IDF, is partially overlapped in the multiple-pump configurations considered in this work as illustrated in Fig.~\ref{fig:setup}(c). On the S--band, besides pumps $P_{6-8}$, there are also contributions of pumps $P_{1-5}$ because the S--band lies within the Raman gain spectrum bandwidth of all these pumps. This makes the design more complex on this region. Thus, although it is expected that the machine learning framework is able to deal with these broadband effects when adjusting the pumps (once the two stages on the S+C+L--band discrete RA are jointly trained), it is also expected to achieve a higher error on the S--band.

It is observed in Fig.~\ref{fig:results_flat_tilted}(h)-(i) that the $E_{MAX}/BW$ for the discrete RA in C+L and S+C+L--band is similar for the flat and the tilted gain profiles. The $E_{MAX}/BW$ is kept below $1.1\cdot10^{-1}$ and $0.9\cdot10^{-1}$~dB/THz for the design of flat and tilted gain profiles, respectively. On the other hand, the $E_{MAX}/BW$ for the distributed RA shown in Fig.~\ref{fig:results_flat_tilted}(g) is higher for the design of the flat gains, but it is still kept below $1.4\cdot10^{-1}$~dB/THz. The reason may be related to the pump distributions, i.e.~the number of pumps and wavelength being more suitable to provide a tilted gain profile. This can be observed on the experimental data--set gain profiles shown in Fig.~\ref{fig:datasets}. The same analysis does not apply for the S+C+L--band, since there it no clear flat/tilted profile trend on its data--set gain curves. Therefore, we also need to take into account that there will be a limitation on the theoretically achievable gain tilt and flatness given experimental set--up that has fixed wavelengths of pump lasers. Fig.~\ref{fig:results_flat_tilted}(j)-(l) shows $RMSE/BW$ and it observed that the trends are very similar to as for $E_{MAX}/BW$.

\vspace{0.15cm}

Finally, we have demonstrated that by only changing the pump powers we are able to achieve low design errors for arbitrary, flat and tilted gain profiles. In conclusion, adjusting the pump powers only, may be sufficient to obtain low errors for various gain profiles. This also points in the direction that the Raman gain profile is more sensitive to pump lasers powers with sufficient number of pump frequencies evenly distributed. We may expect even lower errors if we are able to control pump laser frequencies. However, there are no tunable pump laser available within the considered frequency ranges.

\begin{figure}[!t]
  \centering
  \includegraphics[width=0.47\textwidth]{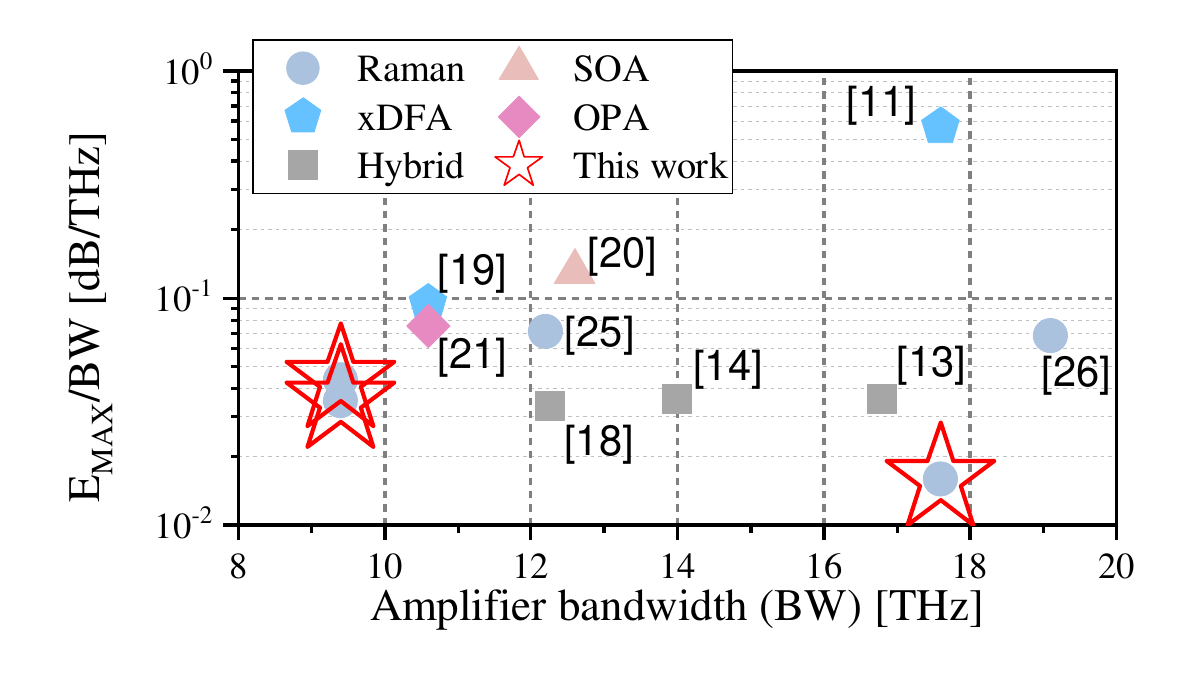}
\caption{$E_{MAX}/BW$ as a function of amplifier bandwidth.}
\label{fig:record}
\end{figure}

To put the presented work in the perspective, in Fig.~\ref{fig:record}, $E_{MAX}/BW$, is plotted for various experimental demonstrations of multi--band amplifiers. It is observed that the presented work results in a low--error and broad bandwidth by means of machine learning.

\section{Conclusion}
\label{sec:conc}
A multi--band programmable gain Raman amplifier operating in C+L and S+C+L--band is experimentally demonstrated. 
The key enabling technique is the machine learning framework that allows for ultra--fast and highly--accurate prediction of the pump currents and voltage for providing the targeted gain profiles. The ability to generate arbitrary gain profiles in a controlled and fast way, may provide novel approaches for the intelligent utilization of the ultra--wideband spectrum and become a key feature for future optical communication systems. Moreover, the programmable gain optical amplifier may advance other areas of fundamental science requiring spectral shaping, such as optical frequency combs.

\section*{Appendix}
The machine learning framework used in this paper to achieve highly accurate Raman amplifier (RA) programmable gains is based on two artificial neural networks. The first neural network $NN_{inv}$ models the RA inverse mapping, i.e. the mapping between gain profiles and pump lasers' currents/voltage. Whereas the forward mapping, i.e., the mapping between the pump lasers' currents/voltage and gain profiles, is learned by a second neural network $NN_{fwd}$. Following, in Section A we describe how these two NNs are trained for the different RA schemes considered in this paper. We also show their prediction accuracy in Section B. Training and validation are performed on disjoint experimental data--sets, whose total number of elements are shown in Table~\ref{tab:datasets}. Section C presents the pump configuration obtained after using $NN_{inv}+NN_{fwd}$ for flat and tilted gain profiles.

\begin{table}[h]
\centering
\caption{\bf Experimental data--set distribution}
\begin{tabular}{cccc}
\hline
RA scheme & C+L dist. & C+L disc. & S+C+L disc. \\
\hline
Training & 3464 & 3000 & 3000 \\
Validation & 2100 & 2600 & 1025 \\
\hline
\end{tabular}
  \label{tab:datasets}
\end{table}

\subsection{Neural networks training}
\label{sec:training}
$NN_{inv}$ is trained using random projection (RP). This training algorithm, also known as extreme learning machine (ELM)~\cite{Huang2011}, initializes the weights of the hidden layers randomly, according to a normal distribution with mean zero and a certain standard deviation $\sigma_{NN_{init}}$, corresponding to NN initialization variance. This random weight assignment is independent from the training data--set and requires a high number of hidden nodes as these weights are kept untrained. The training data--set is used to optimize only the last layer weight by regularized least squares, with a regularization parameter $\lambda$. Since it is performed in a single step, the training time is drastically reduced when compared to standard approaches that updates all the weights in a numerical iterative routine. $NN_{inv}$ models for each RA scheme are shown in Table~\ref{tab:NNinv}, where $f_{act}$ is the nonlinear activation function for all nodes (except the ones on the last layer, which use linear functions), $numHL$ is the number of hidden layers, $numHN$ is the number of hidden nodes, and $D_{input/output}$ is the input/output dimension. To reduce the impact of the randomly initialized weights on the RP method, 20 parallel and independent $NN_{inv}$ are trained and the pump configuration prediction is the average of the 20 $NN_{inv}$ outputs~\cite{Zibar20}. In Table~\ref{tab:NNinv}, $f_{act}$, $numHN$, $\sigma_{NN_{init}}$ and $\lambda$ were obtained after a hyperparameter optimization routine using k-fold cross validation~\cite{bishop2006}.

\begin{table}[b]
\centering
\caption{\bf Neural network models for $NN_{inv}$}
\begin{tabular}{cccc}
\hline
RA scheme & C+L dist. & C+L disc. & S+C+L disc. \\
\hline
Training alg. & RP & RP & RP \\
$f_{act}$ & logsig & sine & sine \\
$numHL$ & 1 & 1 & 1 \\
$numHN$ & 760 & 500 & 500 \\
$D_{input}$ & 90 & 90 & 148 \\
$D_{output}$ & 5 & 5 & 8 \\
$numHN$ & 760 & 500 & 500 \\
$\sigma_{NN_{init}}$ & $6.0 \cdot 10^{-3}$ & $2.6 \cdot 10^{-2}$ & $1.0 \cdot 10^{-2}$ \\
$\lambda$ & $1.0 \cdot 10^{9}$ & $1.0 \cdot 10^{3}$ & $1.0 \cdot 10^{4}$ \\
\hline
\end{tabular}
  \label{tab:NNinv}
\end{table}

$NN_{fwd}$ is trained differently for each RA scheme. For the C+L--band RA (discrete and distributed), $NN_{fwd}$ is trained traditionally updating all weights on the NN iteratively by using the Levenberg-Marquadt (LM) method. However, the high input and output dimensions of the S+C+L--band RA scheme makes the use of LM optimization challenging due to the long convergence time. Thus, RP is applied again only for this scheme. Table~\ref{tab:NNdir} summarizes $NN_{fwd}$ parameters for each RA scheme, where only the RP parameters $f_{act}$, $numHN$, $\sigma_{NN_{init}}$ and $\lambda$ were obtained after a hyperparameter optimization routine. Table~\ref{tab:NNdir} also shows that the RP faster training comes with the cost of having a larger network, with 500 hidden nodes instead of 20 when using LM.

\begin{table}[t]
\centering
\caption{\bf Neural network models for $NN_{fwd}$}
\begin{tabular}{cccc}
\hline
RA scheme & C+L dist. & C+L disc. & S+C+L disc. \\
\hline
Training alg. & LM & LM & RP \\
$f_{act}$ & tanh & tanh & tanh \\
$numHL$ & 2 & 2 & 1 \\
$numHN$ & 10 & 10 & 500 \\
$D_{input}$ & 5 & 5 & 8 \\
$D_{output}$ & 90 & 90 & 148 \\
$\sigma_{NN_{init}}$ & * & * & $1.0 \cdot 10^{-3}$ \\
$\lambda$ & ** & ** & $1.0 \cdot 10^{8}$ \\
\hline
\end{tabular}
\vspace{1ex}
{\\ \raggedright (*) Nguyen-Widrow initialization algorithm~\cite{initnw}; 
    (**) Dynamically modified during training according to~\cite{Hagan94}. \par}
  \label{tab:NNdir}
\end{table}

\subsection{Neural networks validation}
\begin{figure}[t]
  \centering
  \includegraphics[width=0.4\textwidth]{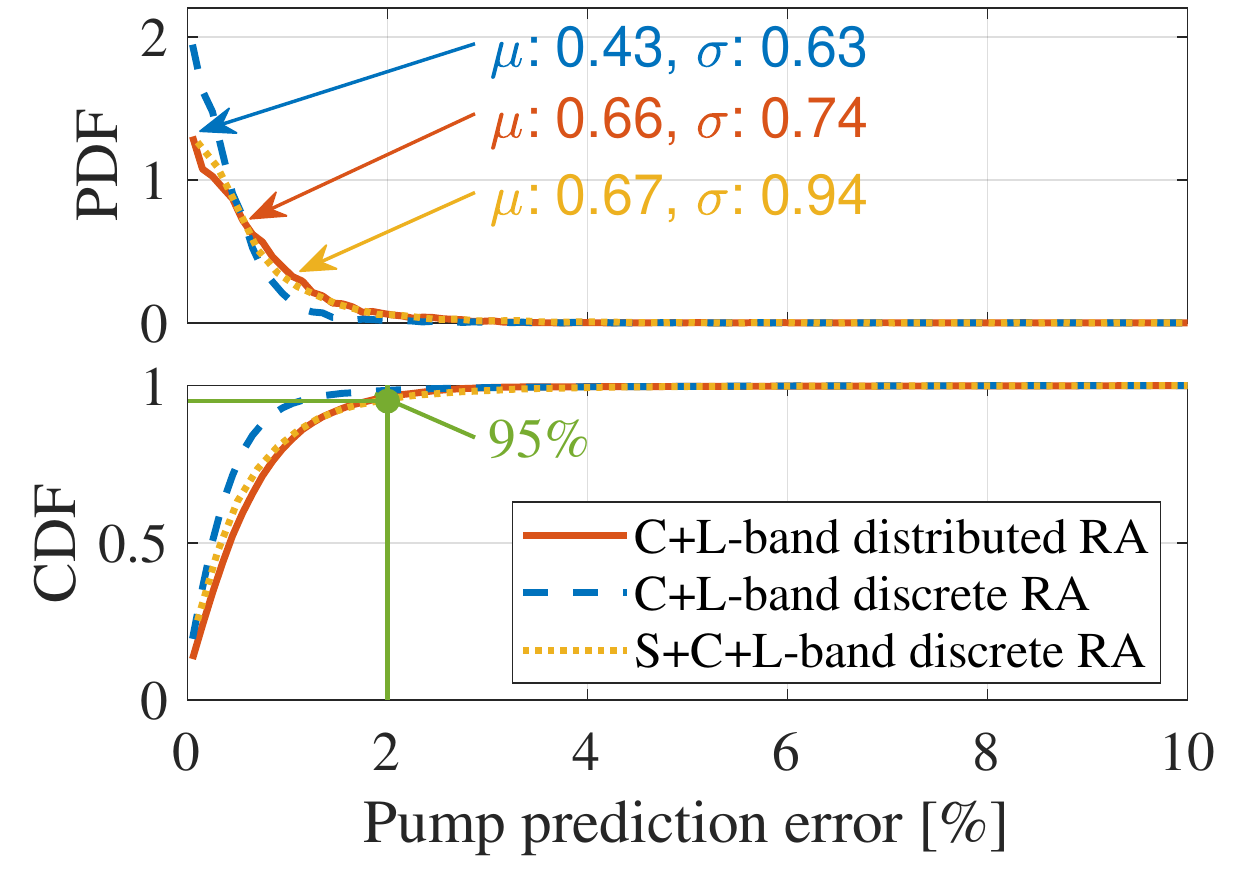}
\caption{Probability density function (PDF) and cumulative density function (CDF) of the $NN_{inv}$ pump current/voltage prediction error, with indication of mean, $\mu$ and standard deviation, $\sigma$.}
\label{fig:NNinv}
\end{figure}

$NN_{inv}$'s performance in predicting pump currents/voltage is presented in Fig.~\ref{fig:NNinv}. The metric used is the absolute error relative to the maximum current/voltage excursion for each pump laser. Fig.~\ref{fig:NNinv} shows the probability density functions (PDF) and the cumulative density functions (CDF) over all the cases on the validation data--set and all pump lasers. Notice that the errors are kept bellow 2\% for 95\% of the cases for all the RA schemes. 

\begin{figure}[t]
  \centering
  \includegraphics[width=0.4\textwidth]{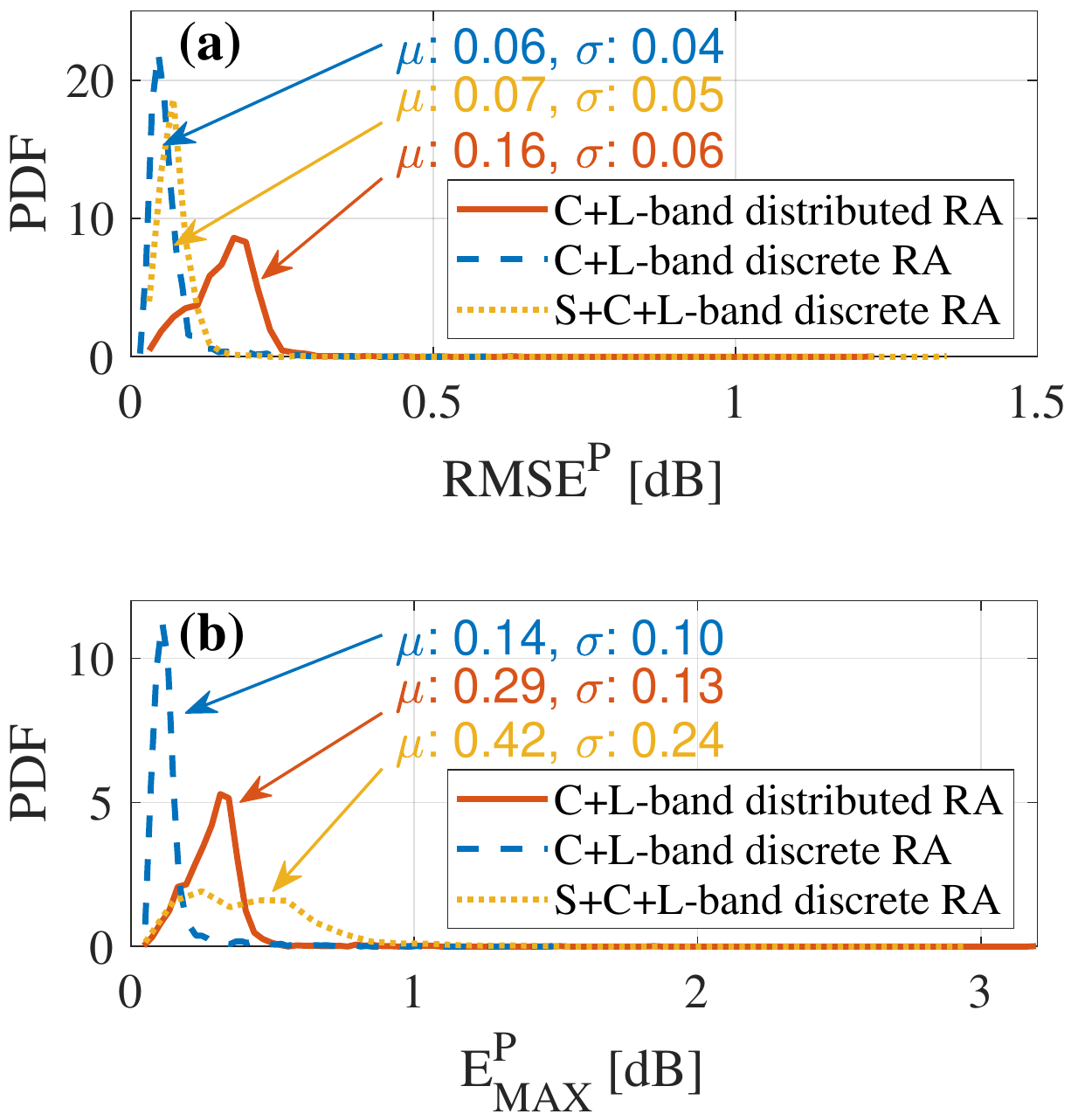}
\caption{Probability density function (PDF) of the $NN_{fwd}$ gain prediction error: (a) $RMSE$ and (b) $E_{MAX}$ with indication of mean, $\mu$, standard deviation, $\sigma$, and maximum (max) values.}
\label{fig:NNdir}
\end{figure}

The prediction performance for the $NN_{fwd}$ is evaluated in terms of root mean squared error ($RMSE$) and maximum absolute error ($E_{MAX}$) between predicted $G_P$ and target $G_T$ gain profiles, extracted from the $K$ WDM points (spectrum), given by
\begin{equation}
RMSE = \sqrt{\frac{1}{K}\sum_{i=1}^{K}(G_{P,i}-G_{T,i})^2},
\label{eq:mse}
\end{equation}
\begin{multline}
E_{MAX} = max\{|G_{P,1}-G_{T,1}|, |G_{P,2}-G_{T,2}|, \cdots \\ 
\cdots, |G_{P,K}-G_{T,K}| \}
\end{multline}
\noindent where $K=90$ and $K=148$ for C+L and S+C+L-band RAs, respectively. Fig.~\ref{fig:NNdir} shows the PDF for $RMSE$ and $E_{MAX}$ over all the cases on the validation data--set.

In Fig.~\ref{fig:NNdir}, the overall $NN_{fwd}$ performances for both C-L--band RAs are consistent with the ones obtained in~\cite{earlyAccessBrusinJLT2020}, which also considers a C+L--band RA (distributed scheme only) with same NN model and training algorithms. On the other hand, the worst performance obtained here by the S+C+L--band RA scheme in terms of $E_{MAX}$ can be explained by its more complex mapping relating more pumps to the gain over a wider bandwidth. S+C+L--band RA scheme was also the only model that used RP, but the same study presented in~\cite{earlyAccessBrusinJLT2020} showed that, for the Raman amplifier case, the performance of the LM only overcomes the RP for higher number of hidden nodes, which requires even more time to train.

The errors $RMSE$ and $E_{MAX}$ are non-convex and unknown functions of the pump configuration that might not share the same local minimums, i.e. the pump configuration that minimizes $RMSE$ might not minimize $E_{MAX}$. However, since the fine--optimization is a gradient-based procedure, it needs to use a differentiable cost function with respect to the pump parameters, which makes the $MSE$ the only candidate for this. When the pdf curves in Fig.~\ref{fig:NNdir}(a) and (b) present similar shapes, like for the C+L--band RAs, it might be an indication that minimums of these two errors occur for similar pump configurations and, consequently, minimizing $MSE$ (which scales with the $RMSE$), may also minimizes $E_{MAX}$. For the S-C-L--band RA, on the other hand, where $E_{MAX}$ and $RMSE$ pdf curves have completely different shapes, it is more likely that minimizing $MSE$ is not the same as minimizing $E_{MAX}$. 

\subsection{Pump configuration for flat and tilted gain profiles}
The pump configurations to achieve flat and tilted gain profiles in Fig.~\ref{fig:results_flat_tilted} are shown in Tables~\ref{tab:pump_flat_CL_dist} to \ref{tab:pump_tilted_SCL}. Current/voltage values are presented from the minimum (first line) to the maximum (last line) gain values, i.e., for the distributed C+L RA case presented in Table~\ref{tab:pump_flat_CL_dist}, the first line corresponds to the bottom gain curve in Fig.~\ref{fig:results_flat_tilted}(a) (minimum gain), and the last line corresponds to the upper gain curve. Recall that these values were obtained after fine optimization.

\begin{table}[!h]
\centering
\caption{\bf {C+L dist. - pump configuration for flat gain profiles}}
\begin{tabular}{ccccc}
\hline
$I_1$ & $I_2$ 	& $I_3$ 		& $I_4$ 	& $I_5$ 	\\
{[A]} & [A] 		& [A] 		& [A] 	& [A] 	\\
\hline
0.43 & 0.20 & 0.23 & 0.42 & 0.22 \\
0.49 & 0.20 & 0.29 & 0.54 & 0.30 \\
0.52 & 0.21 & 0.35 & 0.65 & 0.39 \\
0.50 & 0.22 & 0.40 & 0.74 & 0.53 \\
0.45 & 0.24 & 0.43 & 0.82 & 0.70 \\
0.39 & 0.28 & 0.44 & 0.87 & 0.92 \\
\hline
\end{tabular}
\vspace{1ex}
  \label{tab:pump_flat_CL_dist}
\end{table}

\begin{table}[!h]
\centering
\caption{\bf {C+L dist. - pump configuration for tilted gain profiles}}
\begin{tabular}{ccccc}
\hline
$I_1$ & $I_2$ 	& $I_3$ 		& $I_4$ 	& $I_5$ 	\\
{[A]} & [A] 		& [A] 		& [A] 	& [A] 	\\
\hline
0.49 & 0.20 & 0.21 & 0.22 & 0.20 \\
0.60 & 0.20 & 0.27 & 0.39 & 0.21 \\
0.66 & 0.20 & 0.34 & 0.53 & 0.26 \\
0.67 & 0.21 & 0.41 & 0.66 & 0.35 \\
0.63 & 0.23 & 0.46 & 0.77 & 0.47 \\
0.57 & 0.26 & 0.49 & 0.87 & 0.63 \\
\hline
\end{tabular}
\vspace{1ex}
  \label{tab:pump_tilted_CL_dist}
\end{table}

\begin{table}[!h]
\centering
\caption{\bf {C+L disc. - pump configuration for flat gain profiles}}
\begin{tabular}{ccccc}
\hline
$I_1$ & $I_2$ 	& $I_3$ 		& $I_4$ 	& $I_5$ 	\\
{[A]} & [A] 		& [A] 		& [A] 	& [A] 	\\
\hline
0.51 & 0.26 & 0.27 & 0.45 & 0.29 \\
0.55 & 0.29 & 0.34 & 0.59 & 0.38 \\
0.56 & 0.31 & 0.40 & 0.71 & 0.51 \\
0.52 & 0.35 & 0.45 & 0.82 & 0.68 \\
0.46 & 0.41 & 0.48 & 0.92 & 0.89 \\
\hline
\end{tabular}
\vspace{1ex}
  \label{tab:pump_flat_CL_disc}
\end{table}

\begin{table}[!h]
\centering
\caption{\bf {C+L disc. - pump configuration for tilted gain profiles}}
\begin{tabular}{ccccc}
\hline
$I_1$ & $I_2$ 	& $I_3$ 		& $I_4$ 	& $I_5$ 	\\
{[A]} & [A] 		& [A] 		& [A] 	& [A] 	\\
\hline
0.65 & 0.24 & 0.24 & 0.29 & 0.20 \\
0.71 & 0.30 & 0.32 & 0.45 & 0.23 \\
0.72 & 0.34 & 0.39 & 0.60 & 0.32 \\
0.69 & 0.39 & 0.46 & 0.73 & 0.45 \\
0.63 & 0.44 & 0.50 & 0.85 & 0.63 \\
\hline
\end{tabular}
\vspace{1ex}
  \label{tab:pump_tilted_CL_disc}
\end{table}

\begin{table}[!h]
\centering
\caption{\bf {S+C+L disc. - current and voltage configuration for flat gain profiles}}
\begin{tabular}{cccccccc}
\hline
$I_1$ & $I_2$ & $I_3$ & $I_4$ & $I_5$ & $I_6$ & $I_7$ & $V_8$ \\
{[A]} & [A] & [A] & [A] & [A] & [A] & [A] & [V] \\
\hline
0.72 & 0.47 & 0.49 & 0.72 & 0.61 & 0.37 & 0.60 & 2.01 \\
0.70 & 0.49 & 0.56 & 0.82 & 0.73 & 0.37 & 0.60 & 2.23 \\
0.67 & 0.50 & 0.62 & 0.92 & 0.89 & 0.38 & 0.64 & 2.40 \\
0.65 & 0.49 & 0.65 & 1.07 & 1.09 & 0.41 & 0.85 & 2.40 \\
\hline
\end{tabular}
\vspace{1ex}
  \label{tab:pump_flat_SCL}
\end{table}

\begin{table}[!h]
\centering
\caption{\bf {S+C+L disc. - current and voltage configuration for tilted gain profiles}}
\begin{tabular}{cccccccc}
\hline
$I_1$ & $I_2$ & $I_3$ & $I_4$ & $I_5$ & $I_6$ & $I_7$ & $V_8$ \\
{[A]} & [A] 	& [A] 	& [A] 	& [A] 	& [A] 	& [A] 	& [V]\\
\hline
0.94 & 0.51 & 0.47 & 0.63 & 0.42 & 0.38 & 0.60 & 1.80 \\
0.88 & 0.55 & 0.56 & 0.72 & 0.56 & 0.39 & 0.60 & 1.98 \\
0.85 & 0.57 & 0.65 & 0.82 & 0.69 & 0.39 & 0.60 & 2.19 \\
0.82 & 0.58 & 0.72 & 0.93 & 0.83 & 0.39 & 0.62 & 2.40 \\
\hline
\end{tabular}
\vspace{1ex}
  \label{tab:pump_tilted_SCL}
\end{table}

\section*{Acknowledgment}
This work was supported by the European Union's H2020 program (Marie Sk\l{}odowska-Curie grant 754462 and MSCA-ITN WON grant 814276), the European Research Council (ERC CoG FRECOM grant 771878), the Villum Foundations (VYI OPTIC-AI grant no. 29344), and the UK EPSRC grants EP/M009092/1 and EP/R035342/1.

\ifCLASSOPTIONcaptionsoff
  \newpage
\fi

\bibliographystyle{IEEEtran}
\bibliography{IEEEabrv,bibliography}

\end{document}